\documentstyle[amssymb,aps,prl]{revtex}

\begin{document}
\title{Single particle and collective excitations in the one-dimensional charge
density wave solid\ K$_{0.3}$MoO$_{3}$ \ probed in real time by femtosecond
spectroscopy.}
\author{J.Demsar$^{1}$, K.Biljakovi\'{c}$^{2}$ and D. Mihailovic$^{1}$}
\address{$^{1}$Sol. State Phys. Dept., ''Jozef Stefan'' Institute, Jamova 39, 1001
Ljubljana, Slovenia}
\address{$^{2}$Institute for Physics, Bijeni\v{c}ka 46, HR-10000 Zagreb, Croatia}
\date{\today }
\maketitle

\begin{abstract}
Ultrafast transient reflectivity changes caused by collective and single
particle excitations in the quasi one-dimensional charge-density wave (CDW)\
semiconductor K$_{0.3}$MoO$_{3}$ are investigated with optical pump-probe
spectroscopy. The temperature-dependence of non-equilibrium single particle
excitations across the CDW\ gap and their recombination dynamics are
reported for the first time. In addition, amplitude mode reflectivity
oscillations are observed in real time. A $T$-dependent overdamped response
is also observed which is attributed to relaxation of the phason mode.
\end{abstract}


\twocolumn Molybdenum oxides like K$_{0.3}$MoO$_{3}$ and Rb$_{0.3}$MoO$_{3}$
are well known for their interesting electronic properties arising from
their one-dimensional (1D)\ chain structure\cite{GrunerReview}. At room
temperature they\ are highly anisotropic 1D metals. Upon cooling, they
become susceptible to a Peierls instability on the 1D chains causing
fluctuating local CDW ordering. Upon further cooling, as fluctuations are
reduced, inter-chain interactions cause the CDWs on individual chains to
become correlated, eventually undergoing a second-order phase transition to
a three-dimensionally (3D) ordered state below $T_{c}=183$ K. The formation
of a 3D\ CDW ordered state is concurrent with the appearance of a gap $%
\Delta _{CDW}$ in the single particle (SP) excitation spectrum, while the
collective excitations of the 3D CDW state are described by an amplitudon
mode (AM) and a phase mode (phason).

In this Letter we report femtosecond time-domain transient reflectivity
measurements on K$_{0.3}$MoO$_{3}$ enabling for the first time real-time
observation of the reflectivity modulations caused by collective CDW\
excitations. We report the $T$-dependence of the amplitude $A(T),$ frequency 
$\omega _{A}(T),$ and damping constant $\tau _{A}(T)$ of the AM and for the
first time the $T$-dependence of the phason damping constant $\tau _{p}(T)$%
.\ We also report the $T$-dependence of electron-hole recombination lifetime 
$\tau _{s}$ across the CDW gap below, as well as {\it above} $T_{c}$. Up
till now, enhaced coherent phonon oscillations associated with the formation
of a CDW were observed below $T_{c1}$ in Mo$_{4}$O$_{11}$\cite{Kenji}, but
to our knowledge real-time observation of collective and SP excitations in
CDW\ systems have not yet been reported.

In these experiments, an ultrashort laser {\it pump }pulse first excites
electron-hole pairs via an interband transition in the material (step 1 in
Fig.1a)). In a process which is similar in most materials including metals,
semiconductors and superconductors\cite{Allen,Ippen}, these hot carriers
very rapidly release their energy via $e-e$ and $e-ph$ collisions reaching
states near the Fermi energy within $\tau _{i}=10\sim 100$ fs (step 2 in
Fig.1a)) acting as an ultrashort SP injection pulse. If a CDW or
superconducting gap is present in the SP excitation spectrum, it inhibits
the final relaxation step resulting in a relaxation bottleneck and
photoexcited carriers accumulate above the gap\cite{Kabanov}. This causes a
transient change in reflectivity ${\sl \Delta }{\cal R}/{\cal R}$ due to
change in dielectric constant arising from excited state absorption
processes of the type shown in step 3 in Fig.1a). The density of these
accumulated photoinduced (PI) carriers, $n_{sp}^{\ast }$ can thus be
determined as a function of temperature and time after photoexcitation from
the transient reflectivity change ${\sl \Delta }{\cal R}_{s}/{\cal R}\propto
S(T)e^{-t/\tau _{s}}$, where $\tau _{s}$ is the characteristic SP
recombination time. The amplitude is given by $S(T)\propto n_{sp}^{\ast
}\rho _{2}\left| M_{12}\right| ^{2}$, where $M_{12}$ is the matrix element
for the $E_{1}$ $\rightarrow $ $E_{2}$ optical transition (Fig.1a)) and $%
\rho _{2}$ is the density of states in level $E_{2}$. Whereas $\rho _{2}$
and $M_{12}$ can be assumed to be $T$-independent in first approximation, $%
n_{sp}^{\ast }$ is strongly $T$-dependent when $k_{B}T\sim \Delta _{CDW}.$ A 
$T$-dependence of $n_{sp}^{\ast }$ has recently been calculated for various
gap situations\cite{Kabanov}, which we can now compare with experiments.

In addition to the transient change of reflectivity due to SP excitations
discussed above, a transient reflectivity signal is expected also from
collective modes. The AM is of $A_{1}$ symmetry and involves displacements
of ions about their equilibrium positions $Q_{0}$, which depend on the
instantaneous surrounding electronic density $n(t)$. Since $\tau _{i}<\hbar
/\omega _{A},$ the SP injection pulse may be thought of as a $\delta $%
-function-like perturbation of the charge density $n_{sp}$ and the injection
pulse acts as a time-dependent displacive excitation of the ionic
equilibrium position $Q_{0}(t).$ The response of the AM to this perturbation
is a modulation of the reflectivity\ ${\sl \Delta }{\cal R}_{A}/{\cal R}$ of
the form $A(T)e^{-t/\tau _{A}}\cos (\omega _{A}t+\phi _{0})$ by the
displacive excitation of coherent phonons (DECP) mechanism, known from
femtosecond experiments on semiconductors\cite{Zeiger} and superconductors 
\cite{Kurtz}.

The $\delta $-function-like SP injection pulse also gives rise to the
displacement of charges with respect to the ions, directly exciting the CDW\
phason. Since this is infrared-active, we expect the resulting change of the
dielectric constant $\Delta \epsilon /\epsilon $ to lead to a directly
observable reflectivity transient, which for small $\Delta \epsilon $ can be
approximated as ${\sl \Delta }{\cal R}_{p}/{\cal R}\simeq \Delta \epsilon
/\epsilon .$ In equilibrium, we expect the phason to be pinned and at a
finite frequency, but in non-equilibrium situation such as here, where the
excess carrier kinetic energy may easily exceed the de-pinning energy, the
mode may be de-pinned. In this case we may expect an {\it overdamped}
reflectivity transient written as $P(T)e^{-t/\tau _{p}}\cos (\omega
_{p}t+\phi ),$ with $\omega _{p}\rightarrow 0$, but with a damping constant
which is expected to be similar to that of the AM $\tau _{p}\simeq \tau _{A}$%
, i.e. $\sim $10 ps\cite{Tutis}.

Summing all the contributions, in K$_{0.3}$MoO$_{3}$ the photoinduced
transient reflectivity signal is of the form: 
\begin{eqnarray}
\Delta {\cal R(}t,T{\cal )}/{\cal R} &=&A(T)e^{-t/\tau _{A}}\cos (\omega
_{A}t+\phi _{0})%
\nonumber%
%
\\
&&+P(T)e^{-t/\tau _{p}}+S(T)e^{-t/\tau _{s}}+B(T).
\end{eqnarray}
For completeness we have included an additional term $B(T)$ due to a
long-lived background signal, which is also observed experimentally, and
whose lifetime is longer than the inter-pulse separation of $12$ns. The
different contributions to ${\sl \Delta }{\cal R}/{\cal R}$ can be
effecively distinguished experimentally by their very different
time-dynamics, polarization- and $T$-dependences.

In the experiments reported here a mode-locked Ti:Sapphire laser with
pulselength $\tau _{L}\lesssim 100$ fs at 800 nm was used. The PI change in
reflectivity ${\sl \Delta }{\cal R}/{\cal R}$ was measured using a
photodiode and lock-in detection. The pump laser power was kept below 5 mW,
exciting approximately 10$^{18}$-10$^{19}$ carriers per cm$^{3}$\cite
{carrier density}, and the pump/probe intensity ratio was $\sim $100. The
steady-state heating effect was accounted for as described in Ref.\cite
{ACS98}.\ The experiments were performed on freshly cleaved K$_{0.3}$MoO$%
_{3} $ single crystals with the laser polarization in the {\it a-b} plane, 
{\it a} being the [102] direction and {\it b} is the chain direction\cite
{Xray}. The orientation of the crystal was determined by using an atomic
force microscope$,$ by the direction of the Mo-O chains.

In Fig.1a) we show $\Delta {\cal R}/{\cal R}$ as a function of time at
different temperatures. Below $T_{c}$, an oscillatory component is observed
on top of a negative induced reflection, the latter exhibiting a fast
initial decay followed by a slower decay. As $T_{c}$ is approached from
below, the oscillatory signal dissappears, while the fast transient signal
remains observed well above $T_{c},$ as shown by the trace at 210K. For a
quantitative analysis, we separate the different components of the signal
according to their $T$-dependence and probe polarization anisotropies. In
Fig.1b) we show the signal at $T=45$ and 110K with the oscillatory component
and background $B(T)$ subtracted. The logarithmic plot enables us to clearly
identify two components with substantially different lifetimes, one with $%
\tau _{s}\simeq $ 0.5 ps, and the other with $\tau _{p}\gtrsim 10$ ps at low 
$T.$ Their amplitudes and relaxation times are analyzed by fitting two
exponentials (Eq.(1)). For reasons which will become apparent, we attribute
them to the SP relaxation $S(T)$ and phason relaxation $P(T)$ respectively.
We note that $P(T)$ displays no sign of oscillatory response, in accordance
with the expectation that the phason relaxation is overdamped in this type
of experiment. The insert shows the dependence of the amplitude of the fast
signal on the probe pulse polarization, showing maximum amplitude for $%
\overrightarrow{E}\Vert a$. In Fig.1c) we have plotted only the oscillatory
component with its FFT spectrum, showing a peak at $\nu _{A}=$ 1.7 THz. In
contrast to the transient signal the amplitude of the oscillatory signal is
independent of polarization (Fig.2a)).

The $T$-dependences of the single oscillatory component frequency $\nu _{A}$
and damping $\Gamma _{A}=1/(\pi \tau _{A})$ derived from fits to the
real-time oscillations are shown in Fig.2a). Since the oscillation frequency 
$\nu _{A}$ shows clear softening as T$_{c}$ is approached from below, the
contribution from coherent phonons as observed in Ref.\cite{Kenji} can be
excluded. The measured $\nu _{A}$ and $\Gamma _{A}$ closely follow the
expected behaviour for the AM and are in good agreement with previous
spectroscopic neutron\cite{PougetNeutron} and Raman\cite{WachterRaman} data.
The amplitude of the modulation $A(T)$ falls rather more rapidly with $T$
than $\nu _{A}$ and is rather isotropic in the $a-b$ plane.

In Fig.2b) and c) we have plotted the $T$-dependence of $\tau _{p}$ and $%
P(T) $. At $T=$50 K, $\tau _{p}=12\pm 2$ ps, in agreement with the $\Gamma
=0.05\sim 0.1$ THz linewidths of the pinned phason mode in microwave and IR
experiments\cite{GrunerReview,PougetNeutron,Degiorgi}. With increasing
temperature $\tau _{p}$ is approximately constant up to 90 K and then falls
rapidly as $T\rightarrow T_{c}$. The decrease of $\tau _{p}$ near $T_{c}$ is
consistent with increasing damping due to thermal\ phase fluctuations
arising from coupling with the lattice and SP excitations. The amplitude $%
P(T)$\ exhibits somewhat different behaviour. It appears to first show an
increase with increasing $T$, and then drops as $T\rightarrow T_{c}$. Such $%
T $-dependence behaviour has been previously observed - but not yet
satisfactorily explained - for the threshold field in some non-linear
conductivity experiments\cite{100K}.

Let us now turn to the transient reflectivity signal due to photoexcited SP
excitations. The $T$-dependence of the PI signal amplitude below $T_{c}$ for
a $T$-dependent gap ${\bf \Delta }(T)$ - for simplicity using a BCS-like $T$%
-dependence - is given by\cite{Kabanov}: 
\begin{equation}
S(T)\propto n_{sp}^{\ast }=\frac{{\cal E}_{I}/({\bf \Delta }(T)+k_{B}T/2)}{%
1+\gamma \sqrt{\frac{2k_{B}T}{\pi {\bf \Delta }(T)}}\exp (-{\bf \Delta }%
(T)/k_{B}T)},
\end{equation}
where ${\cal E}_{I}$ is the pump laser intensity per unit cell and $\gamma $
is a constant, depending on the materials' parameters\cite{Kabanov}.
Plotting Eq.(2) as a function of temperature in Fig.3a), we find that the
amplitude $S(T)$ obtained from the fits to the data agrees remarkably well
with the theory for $T<T_{c}$: $S(T)$ is nearly constant up to nearly 100 K,
then increases slightly and then drops very rapidly near $T_{c}$. The value
of the gap $\Delta (0)$=850K$\pm 100$K obtained from the fit of Eq.(2) with $%
\gamma =$ 10 is in good agreement with other measurements \cite{GrunerReview}%
. In contrast to the response of the collective modes $A(T)$ and $P(T),$
both of which dissappear within 10-20 K below $T_{c}$, $S(T)$ remains
observable up to nearly 240 K, i.e. appears to show a pseudogap up to 50 K
above $T_{c}.$ This is - in contrast to the behaviour below $T_{c}$ -
clearly incompatible with a BCS-like description of the gap and suggests the
fluctuating presence of the gap well above $T_{c}$. The polarization
anisotropy of the signal $S(T)$ for $T>T_{c}$ is the same as for $T<T_{c}$
(Fig.1b)), strongly suggesting that the origin of the signal $S(T)$ above $%
T_{c}$ is the same as below $T_{c}$ i.e. SP\ gap excitations.

The $T$-dependence of the relaxation time $\tau _{s}$ (Fig.3b)) is
qualitatively different to $\tau _{p}$ and $\tau _{A}$. As $T\rightarrow
T_{c},$ $\tau _{s}$ appears to {\it diverge} and then drops to $\tau
_{s}\sim 0.25$ ps above $T_{c}$. Such behaviour is in agreement with
expected $T$-dependence of SP\ relaxation across the\ gap. The dominant
recombination mechanism across the gap is phonon emission via phonons whose
energy $\hbar \omega _{p}>2\Delta $\cite{Kabanov}$.$ As the gap closes near $%
T_{c}$, more low-energy phonons become available for reabsorption and the
recombination mechanism becomes less and less efficient. The recombination
lifetime near $T_{c}$ can be shown to be inversely proportional to the gap
as $\tau _{s}\propto 1/\Delta (T)$\cite{Kabanov}. The solid line in Fig.3b)
shows a fit to the data using a BCS-like $T$-dependent gap $\Delta _{BCS}(T)$
with $T_{c}^{3D}$=183 K.

To complete the data analysis, we show in Fig.4 the amplitude of the
slowly-decaying background signal $B(T)$ as a function of $T$. Its anomalous 
$T$-dependence clearly rules out a thermal origin. The lifetime $\tau
>10^{-8}$s deduced from the amplitude of $B(T)$ at ''negative times'' i.e.
from the preceding pulse, strongly suggests excitations involving localized
states. From a fit to an Arrhenius law $B(T)=B_{0}\exp [-E_{a}/k_{B}T],$ we
obtain an activation energy $E_{a}/\Delta (0)\sim 0.6\pm 0.2$, suggesting
that the process involves the excitation of carriers from intra-gap states
into the SP continuum $E_{1}$. As the gap closes, excitations from intra-gap
states to the SP states are no longer possible, explaining the $T$%
-dependence of the signal above $T_{c}$. The microscopic origin of these
states is most likely trapped defects, but the possibility of a collective
excitation cannot be excluded at this stage. We note that a long-lived
signal with a similar $T$-dependence was recently observed in time-resolved
experiments on the cuprate superconductor YBa$_{2}$Cu$_{3}$O$_{7-\delta }$%
\cite{Stevens} and attributed to localized intra-gap states\cite{Slow}.

The real-time optical data presents some qualitatively new information on
the SP and collective excitations in quasi-1D materials. We have found that
because of the qualitatively different time-, polarization- and $T$%
-characteristics, the responses of the different components can be
effectively separated. Apart from directly extracting the $T$-dependences of 
$A(T)$, $\nu _{A}(T)$ and $\tau _{A}(T)$, we also observe an overdamped
mode, which we have assigned to relaxation of the phason mode. In addition
to the observation of the $T$-dependence of photoinduced SP population as
predicted by theory\cite{Kabanov}, we find - also in agreement with
calculations\cite{Kabanov} - that the SP\ recombination time across the gap
diverges as $\tau _{s}\propto 1/\Delta $ as $T\rightarrow T_{c}$. From the
fact that the SP population appears to persist above $T_{c}$, the data shows
clear evidence for the existence of a\ pseudogap for SP excitations above $%
T_{c}$ and suggests the fluctuating presence of a SP{\it \ }gap\cite
{Fluctuations}, rather than fluctuations of the order parameter, which would
appear as a tail also in the SP\ relaxation time $\tau _{s}$ above $T_{c}$ -
but does not. Finally, we should mention that many of the features,
particularly the $T$-dependence of the SP excitation amplitude and the SP\
recombination lifetime is very similar to the behaviour recently reported in
cuprates\cite{Kabanov,Demsar}.\ We note that although coherent oscillations
were reported in YBa$_{2}$Cu$_{3}$O$_{7-\delta }$, no $T$-dependence of the
oscillation frequency was reported, and they were attributed not to any
collective electronic mode, but to $c$-axis phonons\cite{Kurtz}.

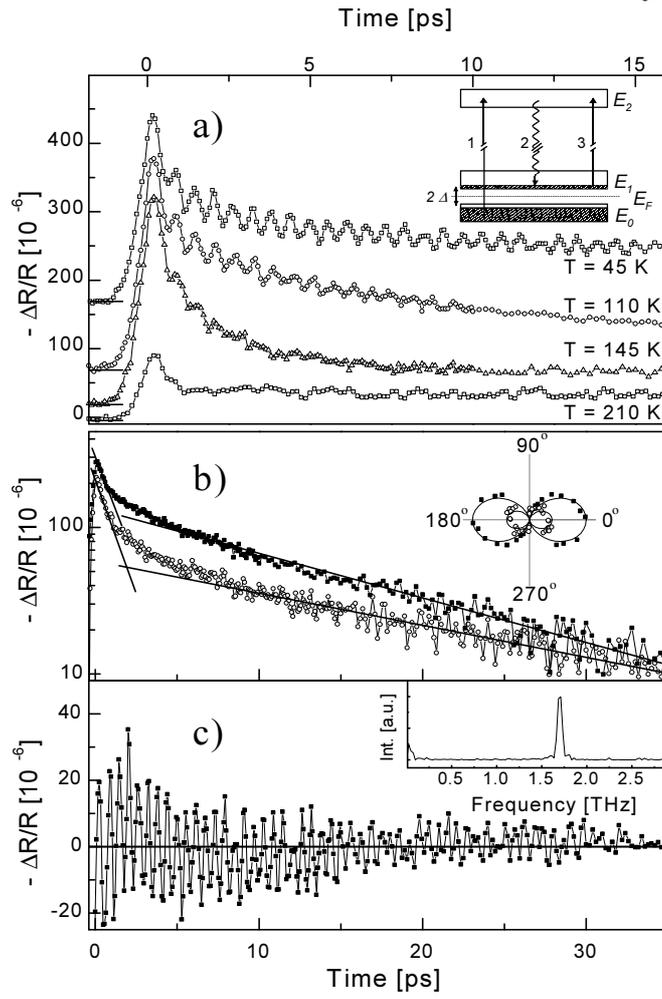
FIG 1. a) The transient reflection $\Delta R/R$ from K$_{0.3}$MoO$_{3}$
after photoexcitation by a 100 fs laser pulse at a number of temperatures
above and below $T_{c}^{3D}=183$ K. The constant background signal $B(T)$
was substracted and offset for clarity. b)\ The time-evolution of transient
signal with the oscillatory component subtracted shown at $T$=45 K and 110
K, displayed on a logarithmic scale to emphasise the difference in
relaxation times $\tau _{p}$ and $\tau _{s}$. The insert shows the amplitude 
$S(T)$ as a function of probe pulse polarization below (solid squares), and
above $T_{c}^{3D}$ (open circles) with respect to the crystal [102] direction%
$.$ c) The oscillatory transient signal $\Delta R_{A}/R$ after subtraction
of the decay components $B$, $S$ and $P$. The fit is made using the first
term in Eq.(1). The insert shows the FFT spectrum of the signal.

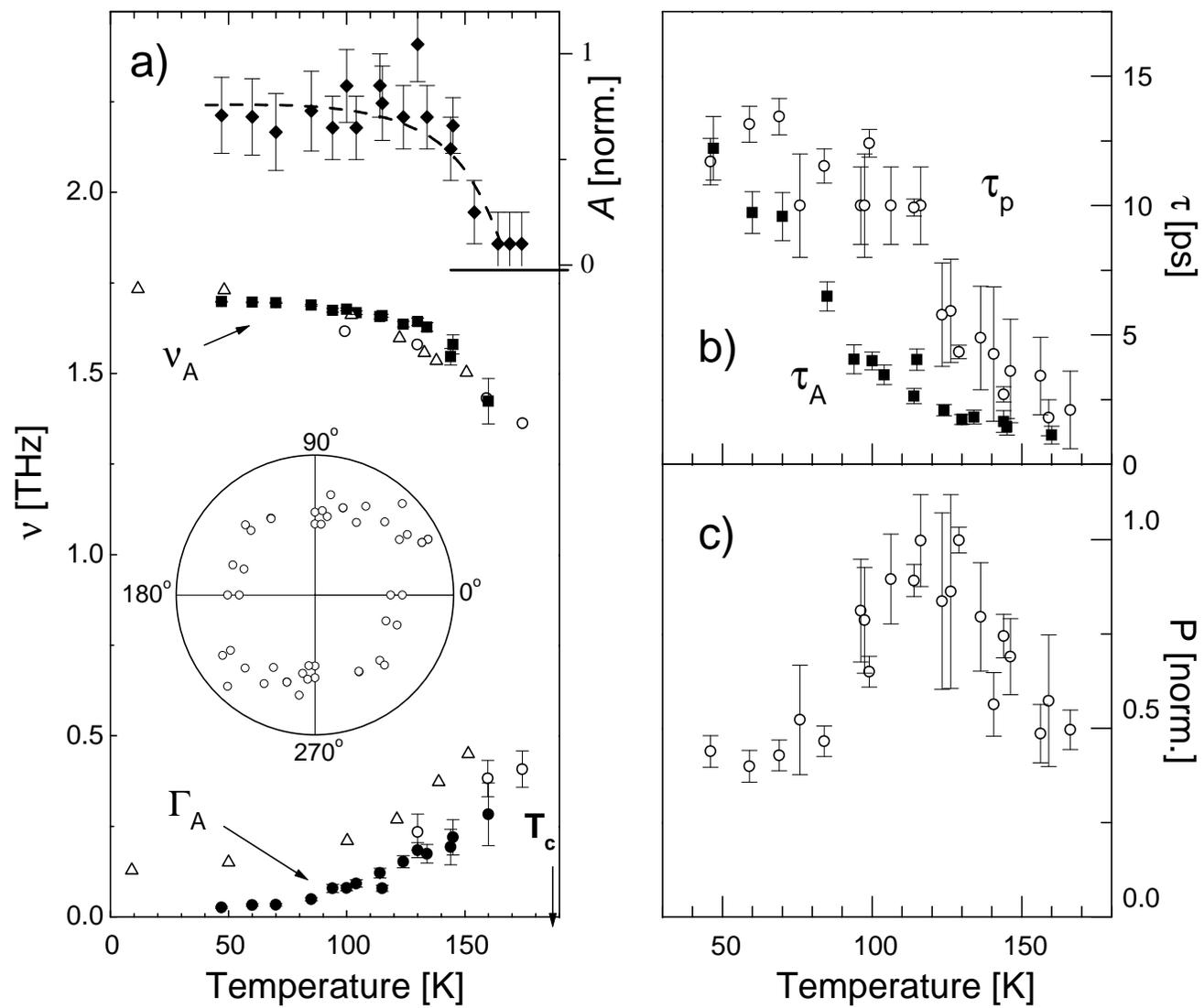
FIG 2. a)\ The amplitude $A(T)$ (diamonds), frequency $\nu _{A}$ (squares)
and damping constant $\Gamma _{A}=1/(\pi \tau _{A})$ (full circles) as a
function of $T$. Data from Refs.\cite{PougetNeutron} (open circles) and \cite
{WachterRaman} (open triangles) are also included. The insert shows the
amplitude $A(T)$ as a function of probe polarization with respect to the
crystal [102] direction. b) $\tau _{p}$ as a function of $T$ (circles). The
amplitudon decay time $\tau _{A}$ is plotted for comparison (squares). c)\ $%
P(T)$ as a function of $T$.

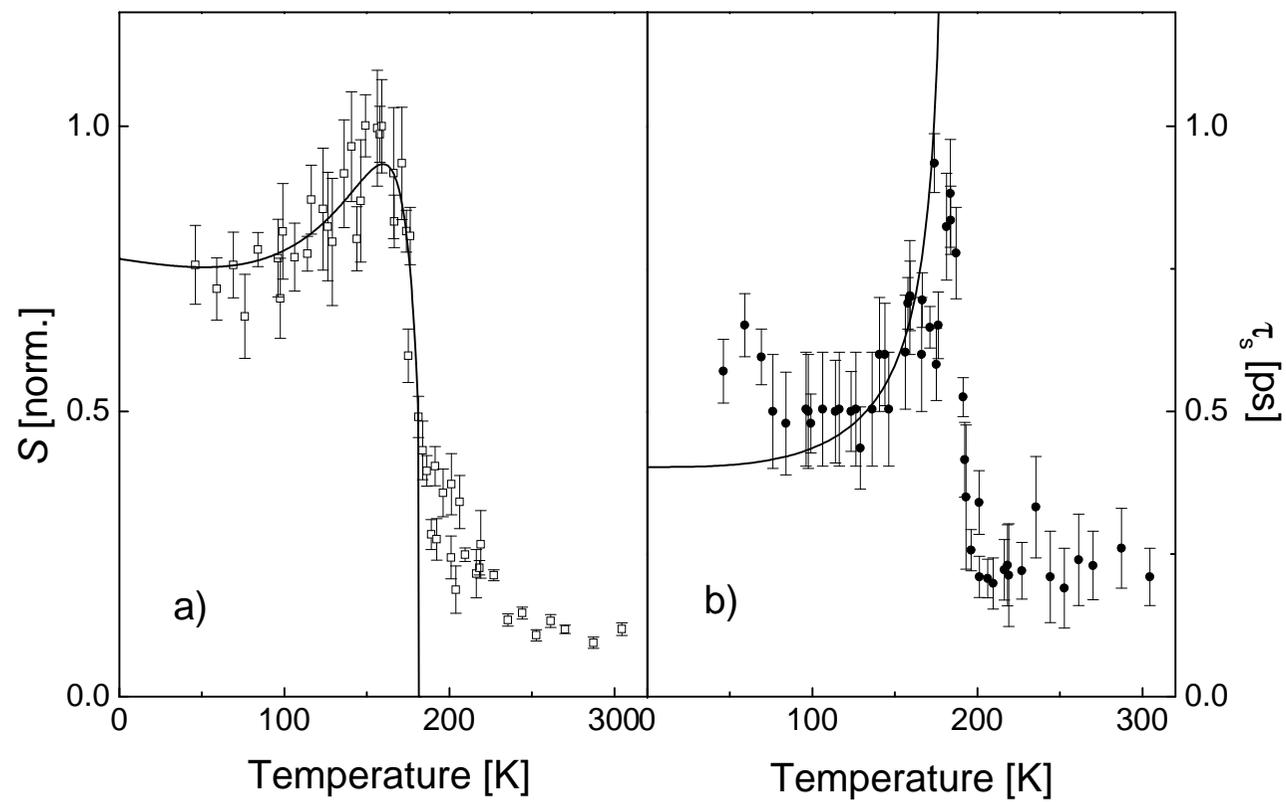
FIG 3. a)\ The $T$-dependence of $S(T)$ . The fit to the data for $S(T)$ is
shown using Eq.(2) with a BCS-like gap $\Delta _{BCS}(T)$ opening at $%
T_{c}^{3D}=183$ K. b) The SP relaxation time $\tau _{s}$ as a function of $T$
and a fit using Eq.(25) of Ref.\cite{Kabanov}.

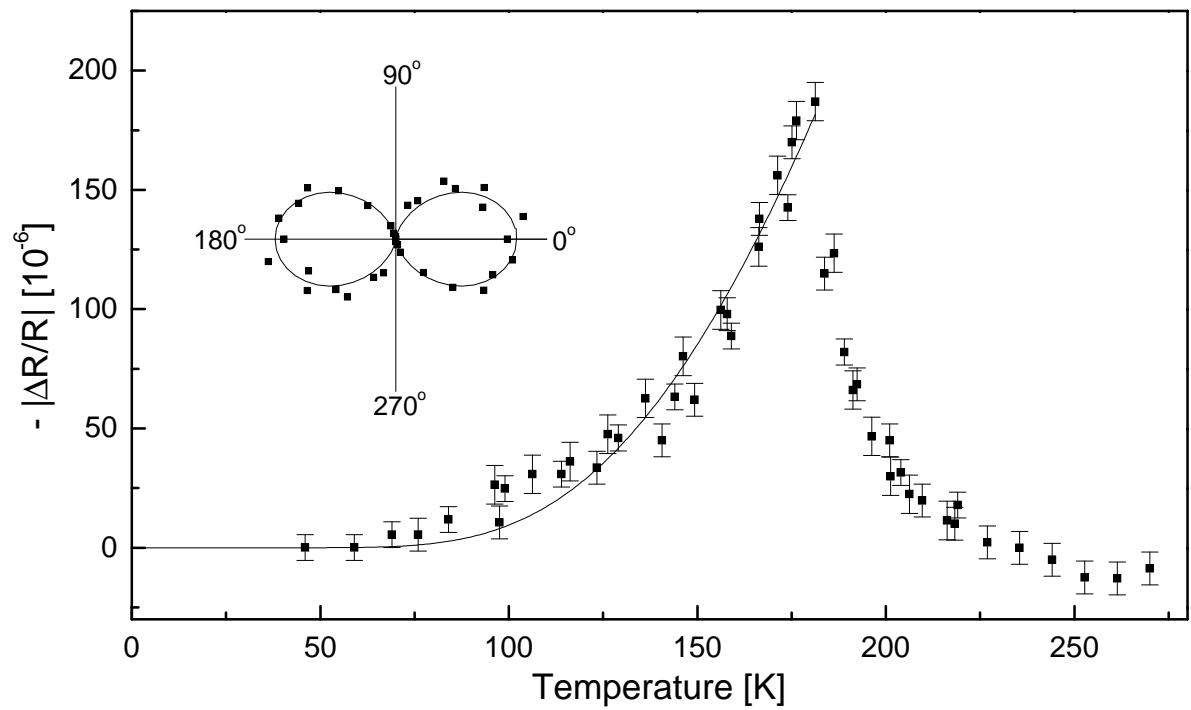
FIG 4. The $T$-dependence of the long-lived signal $B\left( T\right) $. The
line is an Arrhenius fit of the signal amplitude below $T_{c\text{ }}$with $%
E_{a}=60$meV. The insert shows the amplitude $B(T)$ as a function of probe
pulse polarization with respect to the crystal [102] direction.

The authors wish to acknowledge L.Degiorgi and P.Monceau for very useful
discussions and K.Ko\v{c}evar for the AFM analysis. A part of this work was
performed under the auspices of the EU\ ULTRAFAST\ project.


\begin{references}
\bibitem{GrunerReview}  G.Gr\"{u}ner, {\it Rev.Mod.Phys. }{\bf 60}, 1129
(1988), G.Gr\"{u}ner, {\it Density Waves in Solids}, (Addison-Wesley, 1994)

\bibitem{Kenji}  K.Kenji et al., {\it Phys.Rev.B} {\bf 58}, R7484 (1998)

\bibitem{Allen}  P.B.Allen, {\it Phys.Rev.Lett.} {\bf 59}, 1460 (1987)

\bibitem{Ippen}  S. D. Brorson et al., {\it Phys.Rev.Lett.} {\bf 64}, 2172
(1990)

\bibitem{Kabanov}  V.V.Kabanov et al., {\it Phys.Rev. }{\bf B} {\bf 59,}
1497 (1999)

\bibitem{Zeiger}  H.J.Zeiger et al., {\it Phys.Rev.B} {\bf 45}, 768 (1992)

\bibitem{Kurtz}  J.M.Chwalek et al, {\it Appl.Phys.Lett.}{\bf 58}, 980
(1991), W.Albrecht, Th.Kruse and H.Kurtz, {\it Phys.Rev.Lett.} {\bf 69},
1451 (1992), I. I. Mazin et al., {\it Phys.Rev.B} {\bf 49}, 9210 (1994)

\bibitem{Tutis}  E.Tuti\v{s}, S.Bari\v{s}i\'{c}, Phys. Rev.B {\bf 43}, 8431
(1991)

\bibitem{carrier density}  The carrier density was estimated assuming an
absorption length $l=0.1\mu $m at 800 nm.

\bibitem{ACS98}  D. Mihailovic and J.\ Demsar; in {\it Spectrosopy of
Superconducting Materials}, E. Faulques, (ACS Symposium Series 730,
Washington, D.C., 1999)

\bibitem{Xray}  J.Graham and A.D.Wadsley, {\it Acta Cryst. }{\bf 20}, 93,
(1966)

\bibitem{PougetNeutron}  J.P.Pouget et al., {\it Phys. Rev. }{\bf B 43,}
8421 (1991)

\bibitem{WachterRaman}  G.Travaglini, I.M\"{o}rke, P.Wachter, {\it %
Sol.State.Comm. }{\bf 45}, 289 (1983)

\bibitem{Degiorgi}  L.Degiorgi et al., {\it Phys.Rev.B} {\bf 44}, 7808 (1991)

\bibitem{100K}  R.M.Fleming et al.,{\it Phys.Rev.B} {\bf 33}, 5450 (1986),
Dumas et al., {\it Phys.Rev.Lett.} {\bf 50}, 757 (1983), K.Biljakovi\'{c}
(unpublished results)

\bibitem{Demsar}  J.Demsar et al., {\it Phys. Rev. Lett. {\bf 82}, }4918\
(1999)

\bibitem{Stevens}  C.J.Stevens et al, {\it Phys.Rev.Lett.} {\bf 78}, 2212
(1997)

\bibitem{Slow}  V.V.Kabanov, J.Demsar and D.Mihailovic, (to be published,
1999)

\bibitem{Fluctuations}  G.Gr\"{u}ner, {\it Density Waves in Solids},
(Addison-Wesley, 1994), sect. 5.2.
\end{references}
\end{document}